\documentstyle[eqsecnum,aps,psfig,floats]{revtex}
\newcommand{\bR}{{\rm \bf R}}
\newcommand{\bk}{{\rm \bf k}}
\newcommand{\beq}{\begin{equation}}
\newcommand{\eeq}{\end{equation}}
\begin{document}
\draft
\title{The Hubbard model in the two-pole approximation}
\author{A. Avella, F. Mancini\footnote{Corresponding author.  E-mail address: mancini@vaxsa.csied.unisa.it}, D. Villani}
\address{Dipartimento di Scienze Fisiche ``E.R. Caianiello" e Unit\`a I.N.F.M. di Salerno \\
Universit\`a di Salerno, 84081 Baronissi (SA), Italy}
\author{L. Siurakshina, V.Yu. Yushankhai}
\address{LCTA and BLTPh, Joint Institute for Nuclear Research, 141980 Dubna, Russia}
\maketitle
\begin{abstract}
The two-dimensional Hubbard model is analyzed in the framework of the two-pole 
expansion. It is demonstrated that several theoretical approaches, when considered 
at their lowest level, are all equivalent and share the property of satisfying the 
conservation of the first four spectral momenta. It emerges that the various 
methods differ only in the way of fixing the internal parameters and that it exists a 
unique way to preserve simultaneously the Pauli principle and the particle-hole 
symmetry. A comprehensive comparison with respect to some general symmetry 
properties and the data from quantum Monte Carlo analysis shows the relevance of 
imposing the Pauli principle.
\end{abstract}
\pacs
\newpage
\section{Introduction.}
	The Hubbard model [1] plays a key role [2] in the theoretical analysis of highly correlated 
electron systems and in the last years many works have been devoted to its study. The relevance of 
the model is that contains as a main ingredient a competition between the band energy, which tends to 
distribute the electrons throughout the entire crystal, and the repulsive electron-electron interaction 
which tends to localize the electrons on the lattice sites. Although the model is oversimplified and in 
spite of a very intensive study, both analytical and numerical, exact solutions exist only in one and 
infinite dimensions. For finite dimensions greater than one the study is far from being completed and 
more analysis is required. Many techniques and various approximation schemes have been 
formulated; among them we recall the Hubbard I approximation [1], the slave boson method [3-5], 
the non-crossing approximation [6-8], the $d_\infty$ method [9-12], the projection operator method [13-
18], the method of equation of motion [19-21], the spectral density approach [22-24], the composite 
operator method [25-31].

	When the lowest level of approximation is considered, many methods [1, 13-31], though based 
on different schemes, give an equivalent framework of calculation, where the spectral function of the 
single-particle propagator is approximated by a two-pole expansion. A common feature of all these 
methods is that a self-consistent procedure is necessary in order to calculate some parameters which 
appear in the expression of the spectral function. As a consequence, although many methods seems 
to be equivalent, remarkable differences arise, according to the different adopted procedures.

	The purpose of this work is to analyze various methods and compare the different predictions 
with respect to some general symmetry properties of the model and with respect to the numerical data 
obtained by quantum Monte Carlo (qMC) techniques.

\section{The Model.}
The Hubbard model is defined by
\begin{equation}
H=\sum_{ij} t_{ij} c^{\dagger} (i) c(j) +U \sum_i n_{\uparrow}(i) n_\downarrow (i) - \mu \sum_i c^{\dagger} (i)c(i)
\end{equation}     
The first term is a kinetic term that describes the motion of the electrons among the sites of the 
Bravais lattice, defined by the vector set $\{\bR_i\} $. For a two-dimensional squared lattice and by 
restricting the analysis to first nearest neighbors, the hopping matrix $t_{ij}$ has the form
\beq
t_{ij}=-4t\alpha_{ij}=-4t{1\over N} \sum_{\bk} e ^{i\bk\cdot (\bR_i-\bR_j)}\alpha (\bk)	  
\eeq
\beq
\alpha(\bk) = {1\over 2} \left[ \cos (k_x a)+\cos (k_y a)\right]      
\eeq
$a$ being the lattice constant. In addition to the band term, the model contains an interaction term which 
approximates the interaction among the electrons. In the simplest form of the Hubbard model, the 
interaction is between electrons of opposite spin on the same lattice site; the strength of the interaction 
is described by the parameter  $U$. Various generalizations of the model take into account hopping 
between second nearest neighbors and intersite interactions. $c(i)$, $c^{\dagger} (i)$ are annihilation and creation 
operators for $c$-electrons at site $i$  in the spinor notation
\beq
c={c_\uparrow \choose c_\downarrow}\qquad c^\dagger=\left( c^\dagger_\uparrow \quad c^\dagger _\downarrow\right)
\eeq	      
$n_\sigma(i)=c^\dagger_\sigma (i) c_\sigma (i)$ is the number operator of electrons with spin $\sigma=(\uparrow, \downarrow)$
  at the $i$-th site. $\mu$  is the 
chemical potential and is introduced in order to control the band filling $n=\langle c^\dagger (i) c(i)\rangle$. 
By varying 
the ratio  $U/t$, the particle number $n$ and the temperature $T$, it is believed that the Hubbard model is 
capable to describe many properties of strongly correlated fermion systems [2].

	The field $c(i)$ satisfies the equation of motion 
\beq
i{\partial\over \partial t} c(i)=-\mu c(i) + \sum _j t_{ij} c(j) + Un (i) c(i)       
\eeq
where the composite field
\beq
\eta(i) = n(i) c(i)  
\eeq
appears. By means of the Hamiltonian (2.1) one can derive the equation of motion for the new field
\beq
i{\partial\over \partial t} \eta(i)=-(\mu -U) \eta (i) + 4t\pi (i)      
\eeq
where a higher order composite field appears
\beq
\pi(i) = \sum_j \alpha_{ij} \left[ {1\over 2} \sigma^\mu n_\mu (i) c(j) + c(i) c^\dagger (j) c(i)\right]  
\eeq
In Eq. (2.8) we introduced the charge ($\mu=0$) and spin ($\mu=1,2,3$) density operator
\beq
n_\mu(i)=c^\dagger (i) \sigma_\mu c(i)	       
\eeq
with the notation
\beq
\sigma_\mu = (1,\vec \sigma)   \qquad \sigma^\mu = (-1,\vec \sigma)   
\eeq
$\sigma_k$ being the Pauli matrices.

In order to close the infinite hierarchy of equations of motion some truncation is necessary. One 
procedure, common to various methods, is to choose a basis of operators  $\{\Psi(i)\}$ and linearize the 
equation of motion as
\beq
i{\partial\over \partial t} \Psi(i)=\sum _j \varepsilon_{ij}    \Psi(i)    
\eeq	  
where the eigenvalue or energy matrix is self-consistently calculated by means of the equation
\beq
\varepsilon_{ij} = \langle [\Psi (i), H ], \Psi^\dagger (j) ]_{\pm} \rangle /	     
\langle[\Psi (i), \Psi^\dagger (j) ]_{\pm} \rangle                       
\eeq
The symbol $[\ldots ]_\pm$ denotes equal-time anticommutator or commutator, in dependence of the 
statistics of the set  $\{\Psi(i)\}$. The rank of the energy matrix is equal to the number of components of 
$\Psi(i)$. It should be noted that $\varepsilon_{ij}$  may extend over several lattice points, according to the choice of the 
basis $\{\Psi(i)\}$. Different choices for the basis $\{\Psi(i)\}$ can be taken. The mean field approximation 
corresponds to the choice  $\Psi=c$. In Table I we summarize some choices used in various works.
\narrowtext
\begin{table} 
\begin{tabular}{|c|l|}\hline
Refs. 1,20,21 & $\displaystyle\Psi=\left({c \atop \eta}\right)$ \\	      
Ref. 18 & $\displaystyle \Psi=\left({c \atop \eta-\langle n/2 \rangle c}\right)$ \\ 
Refs. 17,26 & $\displaystyle\Psi=\left( { \xi \atop \eta} \right)$ 
\end{tabular}
\caption{\ }
\end{table}
\widetext
The composite operator $\xi(i)$ is defined as 
\beq
\xi(i)	= c(i) - \eta(i) = [ 1-n(i)] c(i)    
\eeq
It is easy to show by direct calculations of the energy matrices that all basis are equivalent and they all 
give place to the same linearized equations of motion. By considering a paramagnetic ground state, 
the thermal retarded Green's function
\beq
S_{\Psi\Psi} (i,j) = \langle R[\Psi (i) \Psi ^\dagger (j) ]\rangle      
\eeq
after the Fourier transformation, has the following expression
\beq
S_{\Psi\Psi} (\bk,\omega)= \sum^2_{i=1} {\sigma^{(i)}_{\Psi\Psi} (\bk) \over \omega - E_i (\bk) + i\eta}
\eeq        
The energy spectra $E_i(\bk)$ are given by 
\beq
E_1 (\bk)=R(\bk) + Q(\bk) \qquad E_2 (\bk)= R(\bk) - Q(\bk)    
\eeq
where
\beq
R(\bk)= {1\over 2} \left[ -2\mu + U\right] - {1\over 2I_1I_2} \left[ m(\bk) + 8t\alpha(\bk)I_1I_2\right]
\eeq	         
\beq
Q(\bk) = {1\over 2} \sqrt{g^2 (\bk) +{4m^2(\bk) \over I_1I_2}}
\eeq	      
and the following notation has been used
$$
I_1=1-n/2 \qquad I_2=n/2 
$$
\beq
m(\bk)=4t\left[ \Delta +\alpha (\bk) (p-I_2)\right]   
\eeq
$$
g(\bk)=-U + {1-n\over I_1 I_2} m(\bk)
$$
The parameters $\Delta$ and $p$ describe a constant shift of the bands and a bandwidth renormalization, 
respectively. They are static intersite correlation functions defined as
\beq
\Delta \equiv \langle \xi^\alpha (i) \xi^\dagger (i) \rangle - \langle \eta ^\alpha (i) \eta^\dagger (i)\rangle    	
\eeq         
\beq
p\equiv {1\over 4} \langle n^\alpha_\mu (i) n_\mu (i) \rangle - \langle \left[ c_\uparrow (i) c_\downarrow (i)
\right]^\alpha c^\dagger_\downarrow(i) c^\dagger_\uparrow (i)\rangle
\eeq      
The notation $c^\alpha(i)$ stands to indicate the field $c$  on the first neighbor sites:
\beq
c^\alpha(i)=\sum_j \alpha_{ij} c(j)
\eeq         
The explicit expressions of the spectral functions $\sigma^{(i)}_{\Psi\Psi} (\bk)$ are given by [26]
$$
\sigma^{(1)}_{\xi\xi} (\bk) = {I_1 \over 2} \left[ 1+ {g(\bk)\over 2Q(\bk) }\right] \qquad  \qquad	
\sigma^{(2)}_{\xi\xi} (\bk) = {I_1 \over 2} \left[ 1- {g(\bk)\over 2Q(\bk) }\right]   
$$
\beq
\sigma^{(1)}_{\xi\eta} (\bk) = {m(\bk)\over 2Q(\bk) }\qquad \qquad \sigma^{(2)}_{\xi\eta} (\bk) = -{m(\bk)\over 2Q(\bk) }
\eeq
$$\sigma^{(1)}_{\eta\eta} (\bk) = {I_2 \over 2} \left[ 1- {g(\bk)\over 2Q(\bk) }\right] \qquad  \qquad	
\sigma^{(2)}_{\eta\eta} (\bk) = {I_2 \over 2} \left[ 1+ {g(\bk)\over 2Q(\bk) }\right]   
$$
The spectral functions relative to other basis can be obtained from (2.23) by recalling that $c(i) = \xi(i) + \eta(i)$.

	Straightforward calculations, reported in Appendix B, show that the first four momenta
\beq
M^{(n)} (\bk) = \int^{+\infty}_{-\infty}	d\omega A (\bk, \omega ) \omega ^n
\eeq                  	
of the electron spectral density
\beq 
A(\bk, \omega) = -{1\over \pi} Im S_{cc} (\bk, \omega)	 = \sum ^2_{i=1} \sigma^{(i)} _{cc} (\bk) \delta
\left[ \omega - E_i(\bk)\right] 
\eeq                
satisfies the exact relations
\beq 
M^{(n)} (\bk)= F.T. \langle \left\{ \left( i{\partial \over \partial t}\right) ^{n-p} c(i),
\left(-i{\partial\over \partial t'}\right)^p c^\dagger (j) \right\}_{E.T.} \rangle\ ,\quad 0\le p\le n
\eeq                 
where $F.T.$ denotes the Fourier transform. Then, there is a complete equivalence between the scheme 
of calculation traced above and the spectral density approach [23], when the two-pole approximation 
is considered.

	Summarizing, when the lowest order is considered all various methods [1, 13-31] are 
equivalent and they all correspond to a two-pole approximation with the same expression for the 
spectral function. The correspondence with the notation of Refs. 20 and 18 is given by
\beq
I_1 I_2 W(\bk) = I_1I_2\left [W_0+ W_1(\bk)\right] = -I_1I_2 \mu - 4 t\Delta + 4t\alpha (\bk)\left[I^2_2 -p\right]
\eeq                   
However, the scheme is not complete unless the self-consistent procedure is defined. The spectral 
functions contain three unknown parameters, $\mu$, $\Delta$, and $p$ very different results are obtained 
according to the methods used to calculate them. This will be discussed in the next Sections.

\section{Self-consistent equations.}
	In Section 2 the energy matrix and therefore the properties of the basic fields $\{\Psi\}$ have been 
fixed by means of Eq. (2.12). However, as we have seen, this constrain does not fix in a unique way 
the dynamics; the parameters $\mu$, $\Delta$, and $p$  remain to be determined and more conditions are required.

	One constrain common to all methods is given by the requirement that the particle number is an 
external parameter, fixed by the boundary conditions. Then, the  following self-consistent equation 
must be satisfied:
\beq 
n=2\left[ 1-C_{cc}\right]
\eeq	      
where the correlation function  $C_{cc}$  is defined in (A.8). We need two more conditions. In Hubbard I 
approximation [1] a simple factorization procedure is used for the two-particle Green's function. 
Then, from the definitions (2.20) and (2.21) one obtains the expressions
\begin{eqnarray}
\Delta &=& 0\\	   
p &=& {n^2\over 4} 	  
\end{eqnarray}
However, Eq. (3.2) is not consistent with the present scheme of calculation. The parameter $\Delta$ is 
directly connected to the single-particle Green's function, and from the definition (2.20) one can 
immediately derive the self-consistent equation
\beq
\Delta = C^\alpha_{\xi\xi} - C^\alpha_{\eta\eta}
\eeq 	 
where the time-independent correlation functions are defined by Eq. (A.6). The parameter p plays an 
important role since it is related to neighboring correlations of the charge, spin and pair. An accurate 
determination of this parameter is very important; as it will be shown in the next Sections, different 
methods of computing p give rise to very different physical solutions.

	In the original work by Roth [20] and in subsequent works [18, 21] the parameter p has been 
calculated by means of the equation of motion in the linearized form (2.11). This procedure leads to 
the following relation
\beq
p=I^2_2 - {(2-\beta \over I_1 (1-\beta^2)} \left[ C^\alpha_{cc} (C^\alpha_{cc}-C^\alpha_{c\eta})+
C^\alpha_{c\eta}	(I^{-1}_2 C^\alpha_{c\eta} -C^\alpha_{cc})\right] - {1\over I_1 I_2 (1-\beta)}
C^\alpha_{c\eta} (C^\alpha_{cc}-C^\alpha_{c\eta})               
\eeq   
where
\beq 
\beta={I_2-C^\alpha_{\eta\eta} - I^2_2 \over I_1I_2}
\eeq	     
and the correlation functions $C^\alpha_{cc}$, $C^\alpha_{c\eta}$ are given by Eq. (A.8).

	In the Composite Operator Method we adopt a different procedure to calculate the parameter $p$. 
This quantity is not expressed in terms of the single-particle propagator, and there is some freedom in 
its determination. In COM we take advantage of this freedom and we fix the parameter $p$ in such a 
way that the Hilbert space has the right properties to conserve the relations among matrix elements 
imposed by symmetry laws.

	There is a wide agreement that the unusual and somehow unexpected properties observed in the 
new class of materials are due to the presence of a high correlation among the electrons. If the 
electron interaction is the key to understand these materials, then it is very crucial that the symmetry 
required by the Pauli principle is correctly treated. A convenient way to take care of the Pauli principle 
is to operate in the representation of second quantization where the Pauli principle manifests through 
the algebra. However, algebra is only one ingredient; physical quantities are expressed in terms of 
expectation values of operators and a suitable choice of the Hilbert space must be made. Physical 
laws, expressed as algebraic relations among the observables, manifest at the level of observation as 
relations among matrix elements.

In a physics dominated by a high correlation among the electrons, we feel that greater attention should 
be dedicated to the conservation of the Pauli principle. It is well known that in most of the 
approximation schemes this symmetry is violated [19], and some alternative schemes should be 
considered. The Pauli principle requires that
\beq
\xi_\sigma (i) \eta^\dagger_\sigma (i) = c_\sigma (i) \left[	n_{-\sigma}(i) - n^2_{-\sigma}(i)\right]
c^\dagger_\sigma (i) = 0 
\eeq             
At level of matrix elements, this condition requires that 
\beq
C_{\xi\eta} = \langle \xi(i) \eta^\dagger (i)\rangle = 0
\eeq	     
By recalling the expression (A.5) for the correlation function $C_{\xi\eta}$   and by means of Eqs. (A.7) and 
(2.19), we easily obtain the following self-consistent equation for the parameter $p$
\beq
p={n\over 2} - \Delta {F_0\over F_1}
\eeq	     
where the quantities $F_0$ and $F_1$ are defined by (A.7).

	Summarizing, the adopted self-consistent procedures are based on the use of Eqs. (3.1), (3.4), 
(3.5) in the Roth's method, and of Eqs. (3.1), (3.4), (3.9) in the COM. It should be noted that the self-
consistent equations are all coupled, so that a different choice for the third equation will have influence 
also on the first two equations. In particular, when the Pauli condition (3.8) is not satisfied, there is an 
ambiguity in writing the first-self consistent equation (3.1). In the next section we shall discuss and 
compare the different results which arise, due to the use of different self-consistent schemes.

\section{The Pauli principle and the particle-hole symmetry.}
	The conservation of the Pauli principle in the Roth's method is studied in Fig. 1, where the 
normalized Pauli amplitude $B_N=\langle \xi (i) \eta^\dagger(i) \rangle /\langle \eta (i) \eta^\dagger (i)\rangle$
is reported vs $n$ for various values of $U$.
\begin{figure}[htb]   
\centerline{\psfig{figure=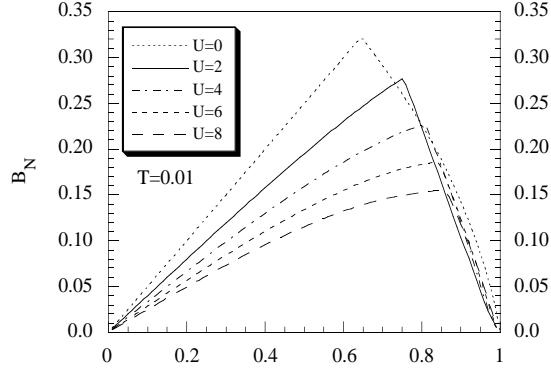,height=6.77cm,width=8.4cm}}
\caption{The normalized Pauli amplitude in the Roth's method  
$B_N=\langle \xi (i) \eta^\dagger(i) \rangle /\langle \eta (i) \eta^\dagger (i)\rangle$ is given as a 
function of the filling for various values of the potential intensity $U$.}
\end{figure}

	We see that in the Roth scheme the Pauli principle is satisfied only in the cases $n=0$ and $n=1$. 
Also, the deviations increase by decreasing U, showing that the non interacting case [i.e. $U=0$] is not 
reproduced in this formulation. At the contrary; the Pauli principle is recovered in the limit  
$U\rightarrow \infty$ for any value of $n$.

	As a consequence of the fact that the Pauli principle is not satisfied also the particle-hole 
symmetry is violated. The Hubbard Hamiltonian has an important property of symmetry: it remains 
invariant under the particle-hole transformation
\beq
c_\sigma (i) \to (-1)^i c^\dagger_\sigma (i)
\eeq	      
By noting that under the transformation (4.1)
\beq
\begin{array}{ll}
c^\dagger c \to 2- c^\dagger c \qquad & c^\dagger \sigma_3 c \to - c^\dagger \sigma_3 c \\
\xi \to \eta^\dagger  & \eta \to \xi ^\dagger    
\end{array}
\eeq    
the self-consistent parameters scale as
\begin{eqnarray}
\mu (2-n) &=& U-\mu (n) \nonumber\\
\Delta (2-n) &=& -\Delta (n) \\
p (2-n) &=& p(n) + (1-n)	\nonumber
\end{eqnarray}        
Then, it is easy to derive the following relations
\beq
\begin{array}{ll}
C_{\xi\xi} (2-n) = C_{\xi\xi} (n) - (1-n) \qquad & C^\alpha _{\xi\xi} (2-n) = C^\alpha_{\eta\eta} (n) \\
C_{\xi\eta} (2-n) = -C_{\xi\eta} (n)   &  C^\alpha _{\xi\eta} (2-n) = C^\alpha_{\xi\eta} (n)  \\
C_{\eta\eta} (2-n) = C_{\eta\eta} (n)   &  C^\alpha _{\eta\eta} (2-n) = C^\alpha_{\xi\xi} (n)  \\
D(2-n) = D(n) + (1-n) & E_s (2-n)= E_s (n)  + U(1-n)   
\end{array}
\eeq      
where $D=\langle n_\uparrow n_\downarrow \rangle $ is the doubly occupancy and  $E_s$ is the energy per site.

On the other hand, by noting that under the transformation $n\to 2-n$ the energy spectra scale as
\beq
\begin{array}{c}
E_1 [2-n,\alpha (\bk)] \to -E_2 [n, -\alpha (\bk)] \\ 
E_2 [2-n,\alpha (\bk)] \to -E_1 [n, -\alpha (\bk)] 
\end{array}
\eeq      
in the two-pole approximation we have [cfr. Appendix A]
\beq
\begin{array}{c}
G_n(2-n) = -(-1)^n G_n (n)  \\
F_n(2-n) = (-1)^n F_n (n)	  \\
B_n(2-n) = -(-1)^n B_n (n)  
\end{array}
\eeq            
Then, it easy to see by direct calculations by means of Eqs. (A.5) and (A.8) that
\beq
\begin{array}{c}
C_{cc} (2-n) = C_{cc} (n) - (1-n) -4C_{\xi\eta} (n) \\
C_{\eta\eta} (2-n) = C_{\eta\eta} (n) + 2C_{\xi\eta} (n)
\end{array}
\eeq         
By comparing (4.4) and (4.7), we see that in the two-pole approximation the particle-hole symmetry 
is conserved if and only if
	\beq
C_{\xi\eta} = \langle \xi(i) \eta^\dagger (i)\rangle = 0
\eeq      
There is a strict relation between the Pauli principle and the particle-hole symmetry. In the two-pole 
approximation the only way to satisfy these symmetry laws is that the parameter $p$ must be fixed in 
such a way that the Pauli principle is conserved.

	In the procedure the parameter p is fixed by means of Eq. (3.5). We have shown in Fig. 1 that 
this choice leads to the result that the Pauli principle is not satisfied. Then, as Eqs. (4.7) show, the 
particle-hole symmetry is violated. This can be seen in Fig. 2a where the 
chemical potential is reported versus $n$ for $T=0.01$ and $U=2,4$. The behavior of $\mu$ does not satisfy the 
law  reported in (4.3). Moreover, as it is more clearly shown in Fig. 2b, in the region $0.7<n<1$ the 
chemical potential has the unphysical behavior to decrease by increasing $n$, which leads to a negative 
compressibility $\kappa={1\over n^2} \left({\partial n \over \partial \mu}\right)$.
\begin{figure}[htb]   
\centerline{\psfig{figure=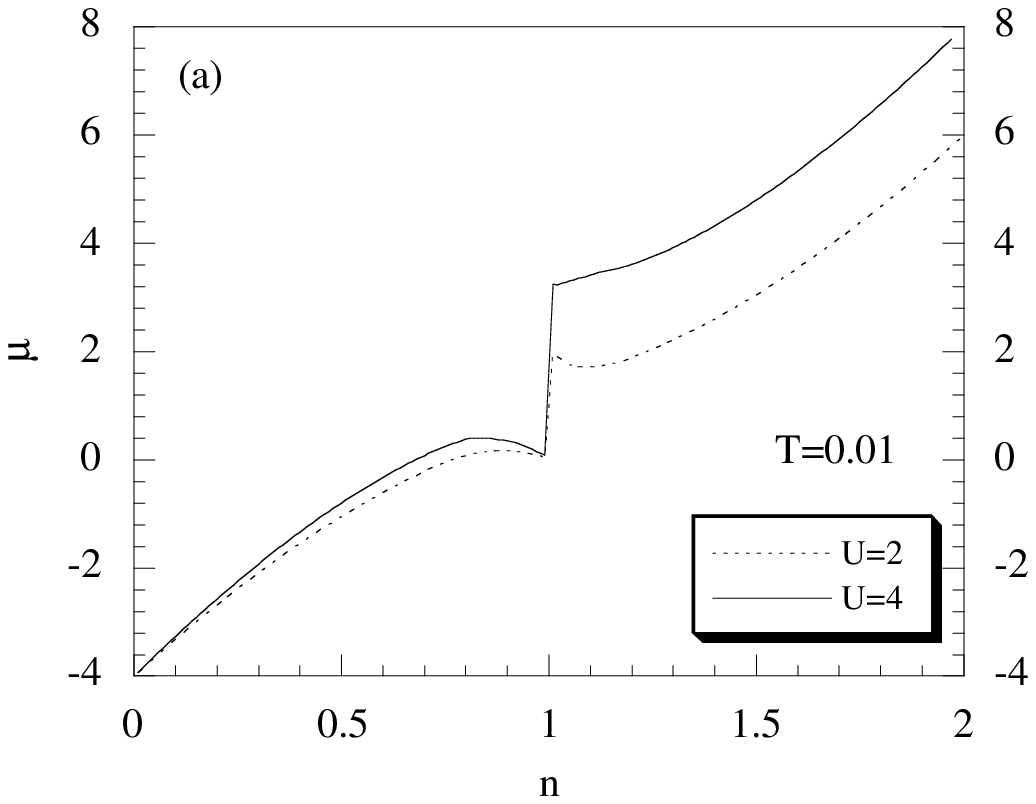,height=6.77cm,width=8.4cm}\psfig{figure=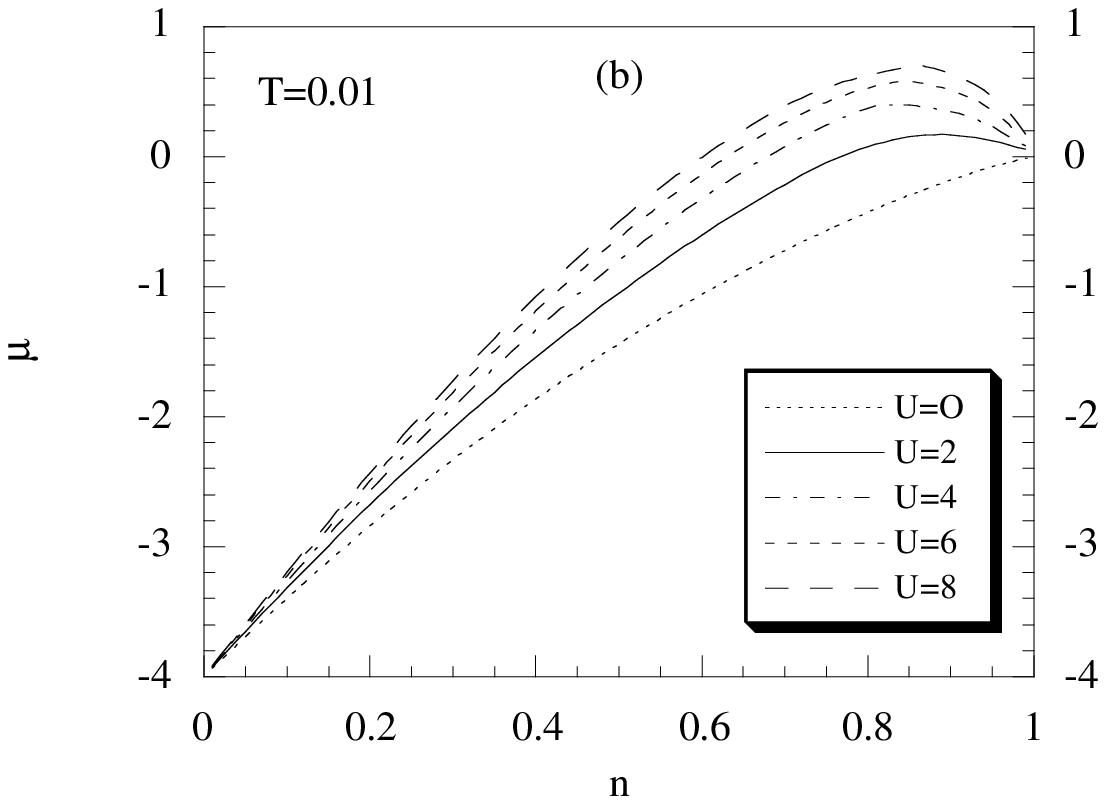,height=6.77cm,width=8.4cm}}
\caption{The chemical potential in the Roth's method is reported as a function of the filling for $T=0.01$ 
and various values of U.}
\end{figure}

\section{COMPARISON WITH MONTE CARLO DATA}
A well developed approach to the study of high correlated electron systems is based on the use 
of numerical techniques. There is now a considerable amount of numerical data on finite-dimension 
lattices; these results offer a precious guide and to them in any case all analytical formulations must 
refer. In this Section we shall present some local quantities, computed in the two-pole approximation. 
In particular, we consider the Roth's method and the COM and compare the results with the 
numerical data by quantum Monte Carlo. In Fig. 3 the chemical potential is given as a function of the 
filling for $U=4$ and $T=0$.
\begin{figure}[htb]
\centerline{\psfig{figure=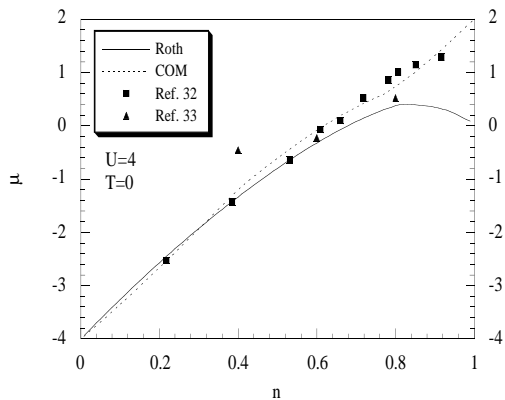,height=6.77cm,width=8.4cm}}
\caption{The chemical potential is given versus the filling $n$ for $T=0^\circ K$ and $U=4$. The solid and 
dotted lines represent the results in the Roth's method and in the COM, respectively. The squares and the 
triangles refer to qMC data and have been taken from Refs. 32 and 33.}
\end{figure}         
The solid line is the result of the Roth procedure; the dotted line is the result of COM. The theoretical data 
are compared with the numerical data by qMC, taken from Refs. 32 and 33. 
\begin{figure}[hbt]
\centerline{\psfig{figure=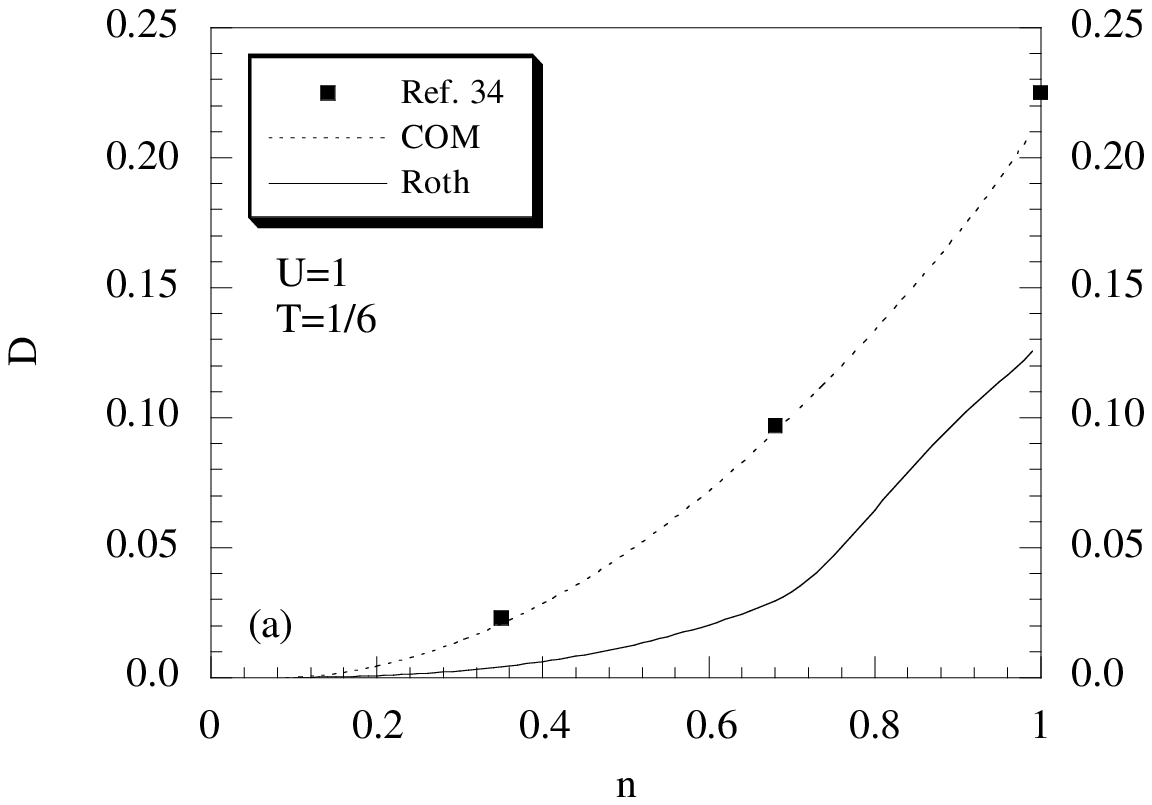,height=6.77cm,width=8.4cm}\psfig{figure=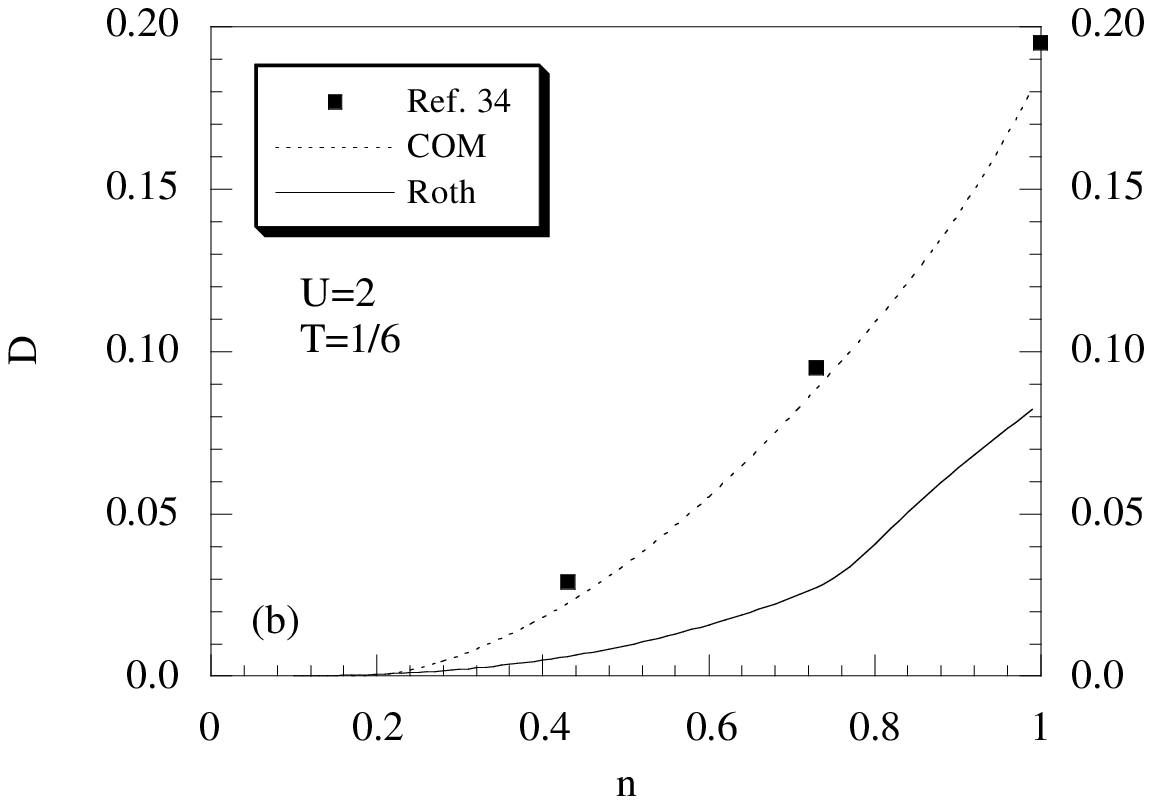,height=6.77cm,width=8.4cm}}
\caption{The double occupancy as a function of n. The notation is the same as in Fig. 3. 
The qMC data are from Ref. 34.}
\end{figure}      
We see that in the region $0.7<n<1$ the Roth results deviate from the behavior predicted by qMC, 
because of the unphysical downward deviation of the chemical potential.

	The double occupancy $D=\langle n_\uparrow n_\downarrow\rangle$ and the energy per site 
$E_s=8t\langle c^\alpha(i) c^\dagger (i)\rangle +UD$
are reported as functions of n in Figs. 4 and 5, respectively. The Roth's and COM's results are compared 
with qMC data, taken from Ref. 34. The parameters have been fixed as $U=1, 2, 4$ and $T=1/6$.
\begin{figure} [hbt]
\centerline{\psfig{figure=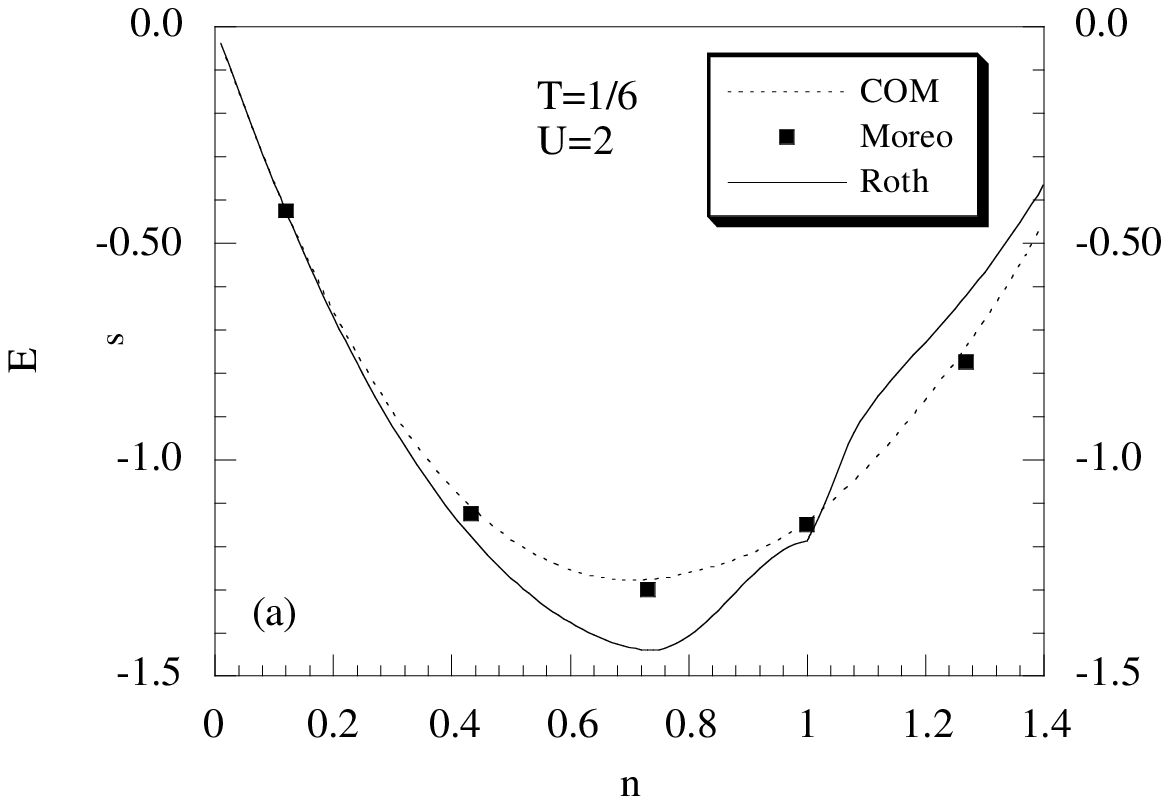,height=6.77cm,width=8.4cm}\psfig{figure=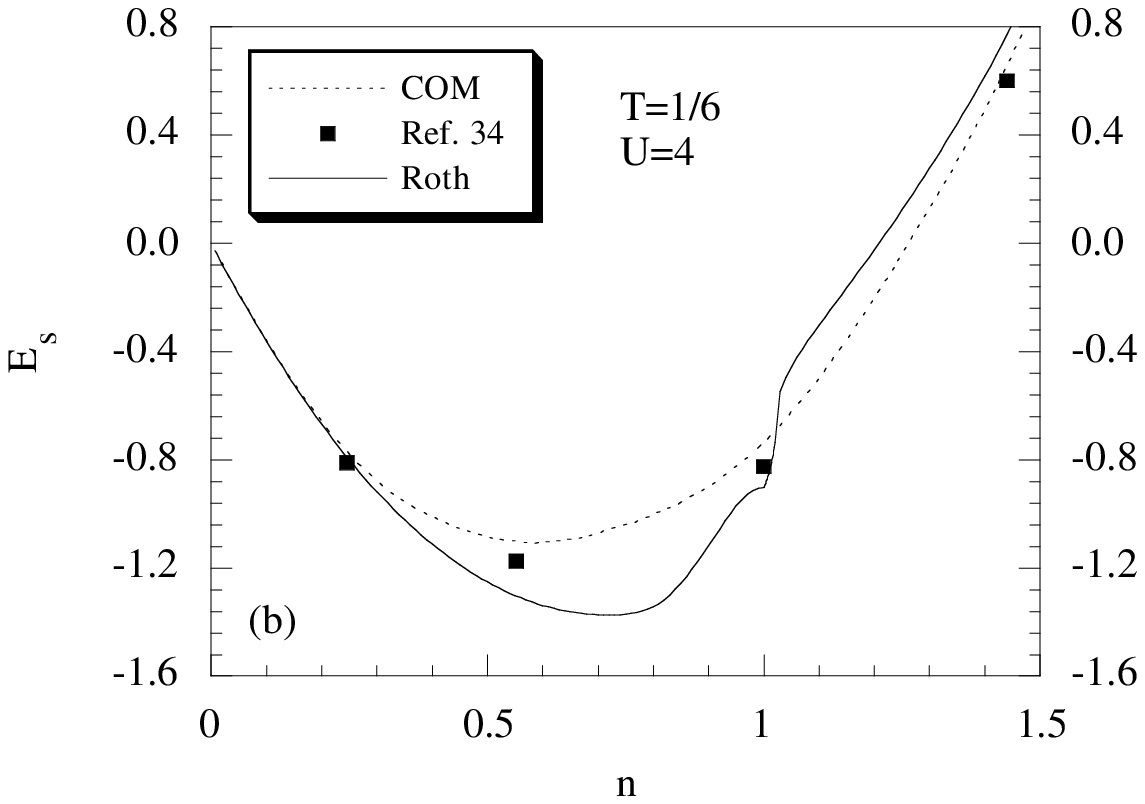,height=6.77cm,width=8.4cm}}
\caption{The energy per site as a function of $n$. The notation is the same as in Fig. 3. The qMC data 
are from Ref. 34.}
\end{figure}     

The results reported in the previous figures show a remarkable difference between the two 
formulations. In the two-pole approximation the method of the equation of motion [19-21] and the 
COM are equivalent, but they differ by the self-consistent procedure used to calculate the parameter p. 
The fact that the symmetry properties are not satisfied in the Roth's scheme leads to the consequence 
that many properties are not conveniently calculated. 

\section{MOTT-HUBBARD TRANSITION.}
	Whether there exists a Mott-Hubbard transition, in the sense that at half filling there is a critical 
value of the Coulomb potential $U_c$ which separates the metallic phase from the insulating phase, is 
another problem still not fully understood in the context of the Hubbard model. In the case of one-
dimensional system it is known that $U_c$ vanishes [35]; in higher dimensions there are no rigorous 
results. It is well known that Hubbard I approximation, defined by Eqs. (3.2)-(3.3), does not predict a 
metal-insulator transition: a gap in the density of states is present for any value of $U\neq 0$. This result 
led Hubbard to improve its approximation [36]. In this Section we shall consider the problem of the 
Mott-Hubbard transition at the light of the formulations discussed above.

	In the case of half filling we have for the energy spectra the expressions
\beq
E_1 (\bk)= R(\bk) + Q(\bk)   \qquad	E_2 (\bk)= R(\bk) - Q(\bk) 
\eeq           
with
\beq 
R(\bk)  = -8 tp\alpha (\bk) \qquad Q(\bk) = {1\over 2} \sqrt {U^2 + [8t(2p-1) \alpha(\bk)]^2}
\eeq	            
It is possible to show that in the Roth scheme for $n=1$ the minimum of $E_1(\bk)$ and the maximum of 
$E_2(\bk)$ are situated at $\bk=(\pi/2, \pi/2)$, so that
\beq 
\Delta E= (E_1)_{\min} - (E_2)_{\max} = U 
\eeq    	        
As in Hubbard I approximation, absence of a metal-insulator transition is found in the Roth's method. 
In COM the minimum of $E_1(\bk)$ is situated at $\bk=(0,0)$, while the maximum of $E_2(\bk)$  is situated at 
$\bk=(\pi,\pi)$, so that
\beq
\Delta E= (E_1)_{\min} - (E_2)_{\max} = -16tp + \sqrt {U^2 + [8t(2p-1) ]^2}        
\eeq                 

The critical value of U which marks the Mott-Hubbard transition is then given by the self-consistent 
equation
\beq
U_c = 8t \sqrt{4p-1}
\eeq        
This equation has been studied in Ref. 37. It has been found that $U_c$ slightly changes with 
temperature. At $T=0^\circ K$, $U_c=1.68 W$, where $W$ is the bandwidth.

\section{CONCLUSIONS}
	We have taken into considerations various methods [1, 13-31] that have been proposed to study 
systems with strong electronic correlations. At the lowest order all these methods coincide, in the 
sense that the spectral function of the single-particle Green's function is approximated by a two-pole 
expansion, and finite life-times effects are neglected. Moreover, the analytic expression for the 
spectral function is the same in all considered methods and is expressed in terms of three self-
consistent parameters. While two parameters, the chemical potential $\mu$ and the static correlation 
function $\Delta \equiv \langle \xi^\alpha (i) \xi^\dagger (i) \rangle - 
\langle \eta ^\alpha (i) \eta^\dagger (i)\rangle$ are controlled by the dynamics through the 
self-consistent equations (3.1) and (3.4), there is some freedom in the determination of the parameter $p$, 
defined by Eq. (2.21), and different choices have been proposed. Most of the works follow the 
procedure suggested by Roth [20], where use has been made of the equation of motion in the 
linearized form (2.11). However, we have shown that this choice violates the Pauli principle and the 
particle-hole symmetry. As a consequence, the solution provides many unpleasant results: the 
compressibility becomes negative in the region of filling $0.7<n<1$, there is absence of a Mott-Hubbard 
transition, the calculated values for some local quantities strongly disagree with the 
numerical data by qMC. We have shown that all these undesirable features disappear when the Pauli 
principle is recovered.

	The matrix energy must be calculated in a self-consistent way, but there is some freedom in 
choosing the procedure. We can take advantage of this freedom to fix the self-consistent procedure in 
such a way that the Hilbert space has the right properties of symmetry. In the framework of two-pole 
approximation the unique way to realize this is to require that the Pauli amplitude (3.8) vanishes.

\appendix
\renewcommand{\theequation}{\thesection.\arabic{equation}}
\section{EQUAL-TIME CORRELATION FUNCTIONS}
	By standard methods equal-time correlation functions can be calculated by means of the 
retarded Green's function. Indeed, one has
 \beq
C_{\Psi\Psi} (\bR_i, \bR_j)\equiv \langle \Psi (\bR_i)\Psi^\dagger (\bR_j)\rangle = {a^2\over (2\pi)^2}
\int d^2 k d \omega [1-f_F (\omega)] \left(-{1\over \pi}\right) Im[ S_{\Psi\Psi} (\bk, \omega)]
\eeq       
where $S_{\Psi\Psi} (\bk, \omega)$ is the Fourier transform of the Green's function. $f_F(\omega)$ is the Fermi 
distribution function and $\Omega_B$ is the first Brillouin zone. By means of Eq. (2.15) we have
\beq
C_{\Psi\Psi} (\bR_i, \bR_j)={\Omega\over 2(2\pi)^2} \sum^2_{i=1} \int _{\Omega_B}d^2 k e^{i{\bk}\cdot (\bR_i-\bR_j)}
[1 + T_i (\bk)]\sigma^{(i)}_{\Psi\Psi} (\bk)
\eeq       
where
\beq
T_i(\bk) = \tanh \left({E_i(\bk) \over 2k_B T}\right)
\eeq     
Let us use the notation
\beq
C_{\Psi\Psi}=C_{\Psi\Psi} (\bR_i, \bR_i)  \qquad C^\alpha_{\Psi\Psi}=C_{\Psi\Psi} (\bR_i, \bR_{i\pm a})  
\eeq    
By using the definition (2.23) for the spectral function we easily obtain the following expressions
\begin{eqnarray}
C_{\xi\xi} &=& \langle \xi (i) \xi^\dagger (i) \rangle = {I_1\over 2}\left[ 1+G_0 - UF_0+{1-n\over {I_1I_2}}
B_0\right] \nonumber \\
C_{\xi\eta} &=& \langle \xi (i) \eta^\dagger (i) \rangle = B_0\\                
C_{\eta\eta} &=& {I_2\over 2}\left[ 1+G_0 + UF_0-{1-n\over I_{1}I_{2}}B_0\right] \nonumber\\
C^\alpha_{\xi\xi} &=& \langle \xi^\alpha (i) \xi^\dagger (i) \rangle = {I_1\over 2}\left[ G_1 - UF_1
+{1-n\over {I_1I_2}}B_1\right]\nonumber\\
C^\alpha_{\xi\eta} &=& \langle \xi^\alpha (i) \eta^\dagger (i) \rangle = B_1\\   
C_{\eta\eta} &=& {I_2\over 2}\left[G_1 + UF_1-{1-n\over I_{1}I_{2}}B_1\right]  \nonumber
\end{eqnarray}     
where we have defined
\begin{eqnarray}
F_n &=& {\Omega\over 2(2\pi)^2} \int_{\Omega_B}d^2 k {[\alpha(\bk)]^n\over 2 Q(\bk)}
[T_1 (\bk) - T_2 (\bk)] \nonumber \\
G_n &=& {\Omega\over 2(2\pi)^2} \int_{\Omega_B}d^2 k [\alpha(\bk)]^n[T_1 (\bk) + T_2 (\bk)] \\                   
B_n &=& {\Omega\over 2(2\pi)^2} \int_{\Omega_B}d^2 k [\alpha(\bk)]^n {m(\bk) \over 2 Q(\bk)} 
[T_1 (\bk) - T_2 (\bk)] \nonumber
\end{eqnarray}
In the text, in order to compare different methods, we are also using the following correlation 
functions
\beq
\begin{array}{ll}
C_{cc}=\langle c(i) c^\dagger(i) \rangle = C_{\xi\xi} + 2C_{\xi\eta} +C_{\eta\eta} \quad &\quad
C_{c\eta}=\langle c(i) \eta^\dagger(i) \rangle = C_{\xi\eta} +C_{\eta\eta} \\
C^\alpha_{cc}=\langle c^\alpha(i) c^\dagger(i) \rangle = C^{\alpha}_{\xi\xi} + 2C^\alpha_{\xi\eta} +C_{\eta\eta} 
\quad &\quad C^\alpha_{c\eta}=\langle c^\alpha (i) \eta^\dagger(i) \rangle = C^\alpha_{\xi\eta} +C^\alpha_{\eta\eta} \\
\end{array}
\eeq        

\section{CONSERVATION OF THE FIRST FOUR SPECTRAL MOMENTA}
Let us show that the first four momenta of the electron spectral density satisfy the exact relations given by 
Eq. (2.26). By straightforward calculations we can see that:
\begin{eqnarray}
M^{(0)} (\bk) &=& F.T. \langle\left\{ c(i), c^\dagger (j) \right\} \rangle = 1	   \\    
M^{(1)} (\bk) &=& F.T. \langle\left\{ \left(i{\partial \over \partial t}\right) c(i), c^\dagger (j) \right\} \rangle = 
-\mu + {n\over 2} U - 4t\alpha (\bk) \\     
M^{(2)} (\bk) &=& F.T. \langle\left\{ \left(i{\partial \over \partial t}\right)^2 c(i), c^\dagger (j) \right\} \rangle = 
-\mu M^{(1)} (\bk) + U\left[ m(\bk) - (\mu-U) {n\over 2} - 4t(\Delta + \alpha (\bk) p)\right] \nonumber \\
&& - 4 t F.T.\sum_l \alpha_{il} \langle\left\{ \left( i{\partial\over \partial t}\right) c(l), 
c^\dagger(j)\right\}\rangle=\nonumber\\ 
&=& \mu^2 - \mu n U +{n\over 2} U^2 +16 t^2\alpha^2 (\bk) +8t\mu \alpha(\bk)- 8t {n\over 2} U \alpha (\bk)\\    
M^{(3)} (\bk) &=& F.T. \langle\left\{ \left(i{\partial \over \partial t}\right)^2 c(i), \left(-i{\partial \over \partial t'}\right)
c^\dagger (j) \right\} \rangle \nonumber \\
&= &-\mu M^{(2)} (\bk) + F.T. \langle\left\{ - \mu \left(i{\partial \over \partial t}\right)
c(i)+ U\left(i{\partial \over \partial t}\right)\eta(i) - 4t \sum _l \alpha_{il} \left(i{\partial \over \partial t}\right)c(l),
U\eta^\dagger (j) - 4t \sum _m \alpha _{mj} c^\dagger (m)\right\}\rangle \nonumber\\
& =& \left[ - 4t\alpha (\bk) - \mu \right]^3 + 3{n\over 2} U \left[ - 4t\alpha (\bk) - \mu \right]^2 + {n\over 2}U^3 -3{n\over 2} 
\mu U^2 -8t{n\over 2} U^2 \alpha (\bk) - 4t\Delta U^2 - 4tpU^2\alpha(k) 
\end{eqnarray}           
On the other hand, from Eqs. 2.23 - 2.25 we have
\begin{eqnarray}
M^{(0)} (\bk) &=& \sum^2_{i=1} \left[\sigma^{(i)}_{\xi\xi} (\bk) + 2\sigma^{(i)}_{\xi\eta} (\bk) + \sigma^{(i)}_{\eta\eta} (\bk)
\right] = 1 \\	          
M^{(1)} (\bk) &=& \sum^2_{i=1} \left[\sigma^{(i)}_{\xi\xi} (\bk) + 2\sigma^{(i)}_{\xi\eta} (\bk) + \sigma^{(i)}_{\eta\eta} (\bk)
\right] E_i(\bk)  \nonumber \\	 
&=& R(\bk) M^{(0)} (\bk) + Q(\bk) \left[{I_1-I_2\over 2} {g(\bk)\over Q(\bk)} + 2{m(\bk)\over Q(\bk)}\right] 
=-\mu +{n\over 2}U - 4t\alpha (\bk)           
\end{eqnarray}
From Eq. (B.6) follows the identity
\beq
{I_1-I_2\over 2} {g(\bk)\over Q(\bk)} + 2{m(\bk)\over Q(\bk)} = -{R(\bk)\over Q(\bk)} +
{M^{(1)} (\bk) \over Q(\bk)}       
\eeq           
that leads to
\begin{eqnarray}
M^{(2)} (\bk) &=& \sum^2_{i=1} \left[\sigma^{(i)}_{\xi\xi} (\bk) + 2\sigma^{(i)}_{\xi\eta} (\bk) + \sigma^{(i)}_{\eta\eta} (\bk)
\right] E^2_i(\bk) \nonumber \\
&=&  \left[R^2(\bk) + Q^2(\bk) \right] M^{(0)} (\bk) +2R(\bk) Q(\bk)\left[ -{R(\bk)\over Q(\bk)} + {M^{(1)}(\bk)\over Q(\bk)}
\right]  \nonumber\\
&=& \mu^2 -\mu nU + {n\over 2}U^2 + 16t^2\alpha^2 (\bk)  + 8t\mu\alpha (\bk) - 8t{n\over2} U \alpha (\bk)   
 \end{eqnarray}	                                   
\begin{eqnarray}
M^{(3)} (\bk) &=& \sum^2_{i=1} \left[\sigma^{(i)}_{\xi\xi} (\bk) + 2\sigma^{(i)}_{\xi\eta} (\bk) + \sigma^{(i)}_{\eta\eta} (\bk)
\right] E^3_i(\bk) \nonumber \\
&=&  \left[R^3(\bk) + 3R(\bk) Q^2(\bk) \right] M^{(0)} (\bk) +\left[ Q^3(\bk) +3R^2(\bk) Q(\bk)\right] 
\left[- {R(\bk)\over Q(\bk)} + {M^{(1)}(\bk)\over Q(\bk)}\right]  \nonumber\\
&=& 2R(\bk) M^{(2)} (\bk) +\left[ Q^2 (\bk) - R^2(\bk)\right] M^{(1)} (\bk) \nonumber \\
& =& \left[ - 4t\alpha (\bk) - \mu \right]^3 + 3{n\over 2} U \left[ - 4t\alpha (\bk) - \mu \right]^2 + {n\over 2}U^3 
-3{n\over 2} \mu U^2 -8t{n\over 2} U^2 \alpha (\bk) - 4t\Delta U^2 - 4tpU^2\alpha(k) 
\end{eqnarray}  
Then, there is a complete equivalence among all considered theoretical approaches, when two-pole 
approximation is used.

\end{document}